\documentclass{naturefig}

\usepackage{color}

\usepackage{amsmath,amsfonts,amssymb}
\usepackage{graphicx}

\title{Energy Scaling of Targeted Optimal Control of Complex Networks}

\author{Isaac Klickstein$^1$, Afroza Shirin$^1$ \& Francesco Sorrentino$^1$}

\begin{document}

\maketitle

\begin{affiliations}
 \item Department of Mechanical Engineering, The University of New Mexico, Albuquerque, NM 87131
\end{affiliations}
%
% Abstract
\begin{abstract}
Recently it has been shown that the control energy required to control a dynamical complex network is prohibitively large when there are only a few control inputs.
Most methods to reduce the control energy have focused on where, in the network, to place additional control inputs.
Here, in contrast, we show that by controlling the states of a subset of the nodes of a network, rather than the state of every node, while holding the number of control signals constant, the required energy to control a portion of the network can be reduced substantially.
The energy requirements exponentially decay with the number of target nodes, suggesting that large networks can be controlled by a relatively small number of inputs as long as the target set is appropriately sized.
We validate our conclusions in model and real networks to arrive at an energy scaling law to better design control objectives regardless of system size, energy restrictions, state restrictions, input node choices and target node choices.
\end{abstract}
%
% Introduction
\indent Recent years have witnessed increased interest from the scientific community regarding the control of complex dynamical networks \cite{sorrentino2007controllability, mikhailov2008introduction,
yu2013synchronization, tang2012evolutionary, wang2002pinning, liu2011controllability, liu2011reply, ruths2014control, summers2014optimal, wang2012control, nepusz2012controlling, yuan2013exact, muller2011few, iudice2015structural}.
Some common types of networks examined throughout the literature are power grids \cite{arianos2009power, pagani2013power}, communication networks \cite{onnela2007analysis, kwak2010twitter}, gene regulatory networks
\cite{
palsson2015systems},
neuronal systems
\cite{
sporns2013structure,
papo2014complex},
food webs
\cite{
allhoff2013evolutionary},
and social systems
\cite{
lerman2010information}.
We define networks as being composed of two components; the nodes which constitute the individual members of the network and the edges which describe the coupling or information sharing between nodes
\cite{
newman2010networks}.
Particular focus has been paid to our ability to control these networks
\cite{
liu2011controllability,
ruths2014control,
summers2014optimal,
wang2012control,
nepusz2012controlling,
yuan2013exact,
iudice2015structural,
gao2016control}.
A network is deemed controllable if a set of appropriate control signals can drive the network from an arbitrary initial condition to any final condition in finite time.
If a network is controllable, a control signal which achieves such a goal is not necessarily unique.\\
%
%% Background Information on Minimum Energy Control
%
\indent One important metric to characterize these control signals is the energy that each one requires.
From optimal control theory, we can define the control action which, for a given distribution of the control input signals satisfies both our initial and final conditions as well as minimizes the energy required to perform the task
\cite{
kailath1980linear}.
The energy associated with the minimum energy control action provides an energetic theoretical limit.
Knowledge of the minimum control energy is crucial in order to understand how expensive it can be to control a given network when applying any alternative control signal.
The minimum energy framework has recently been examined in
\cite{
yan2015spectrum,
yan2012controlling}
which have shown that based on the underlying network structure, the distribution of the control input signals, the desired final state, and other parameters, the energy to control a network may lie on a distribution that spans a broad range of orders of magnitude.
%
%% Targeting 100% of Nodes
%
In this paper, we focus on reducing the energy that is maximum with respect to the choice of the initial state, final state, and, in general, of an arbitrary control action.
We note that in real applications involving large complex networks, achieving control over all of the network nodes is often unfeasible \cite{yan2015spectrum,yan2012controlling} and ultimately unnecessary.\\
%
%% Energy Reduction Techniques
%
\indent One possible method to reduce the required energy was investigated in
\cite{
chen2016energy},
where additional control signals were added in optimal locations in the network according to each node's distance from the current set of control signals.
In \cite{nacher2016minimum,wuchty2014controllability}, the minimum dominating set (MDS) of the underlying graph of a network is determined and each node in the MDS is assumed capable of generating an independent signal along each of its outgoing edges.
As every node not in the MDS is only one edge away from a node in the MDS, and each edge from an MDS node to a non-MDS represents a unique control signal, the control energy will be relatively small.
In this paper, for the first time, we adjust the control goal to affect only a subset of the network nodes, chosen as the targets of the control action, and consider the effect of this choice on the required control energy.
This type of target control action is typically what is needed in applications in gene regulatory networks\\
\cite{
yang2008finding},
financial networks
\cite{
galbiati2013power},
and social systems
\cite{
klemm2003nonequilibrium}.\\
%
%% How does the Energy Scale?
%
Our first main contribution is determining how the energy scales with the cardinality of the target set.
In particular, we find that the minimum control energy to control a portion of the complex network decays exponentially as the number of targets is decreased.
Previous work
\cite{
yan2015spectrum,
yan2012controlling}
has only investigated the control energy for complex networks when the target set coincides with the set of all nodes.
We also look at the energetic relation between the number of targets and other network parameters such as the number of inputs and the amount of time allocated for the control action.
%
%% Does Target Control Benefit Other Control Methods
%
Our second main contribution is showing that target control is applicable to other control actions generated with respect to other cost functions.
Target control has received recent attention in
\cite{
gao2014target,
iudice2015structural}
which examined methods to choose a minimal set of independent control signals necessary to control just the targets.
Here, a target control signal is examined that is optimal with respect to a general quadratic cost function that appears often in the control of many real systems.
%
%% End of introduction
%
%% Begin of Results
\section*{Results}
% 1) Problem Formulation
% 1.a Description of Networks (Nodes and Edges) 
\paragraph{Problem Formulation}
Complex networks consist of two parts; a set of nodes with their interconnections that represent the topology of the network, and the dynamics which describe the time evolution of the network nodes.
First, we summarize the definitions needed to describe a network.
We define $\mathcal{V} = \{i\}$, $i = 1, \ldots, n$ to be the set of $n$ nodes that constitute a network.
The adjacency matrix is a real, square $n \times n$ marix, $A$, which has nonzero elements $A_{ij}$ if node $i$ receives a signal from node $j$.
For each node $i$ we count the number of receiving connections, called the in-degree $k_i^{\text{in}}$ and the number of outgoing connections, called the out-degree $k_i^{\text{out}}$.
The average in-degree and average out-degree for a network is $k_{\text{av}}$.
One common way to characterize the topology of a network is by its degree distribution.
Often the in-degree and out-degree distributions of networks that appear in science and engineering applications are scale-free, i.e., $p(k) \sim k^{-\gamma}$ where $k$ is either the in-degree or out-degree with corresponding $\gamma_{\text{in}}$ and $\gamma_{\text{out}}$, and most often $2 \leq \gamma \leq 3$ \cite{albert2002statistical}.\\
%
% 1.b Description of Linear Dynamics
\indent While most dynamical networks that arise in science and engineering are governed by nonlinear differential equations, the fundamental differences between individual networks and the uncertainty of precise dynamics make any substantial overarching conclusions difficult
\cite{
gao2014target,
albert2002statistical,
liu2011controllability}.
Nonetheless, linear controllers have proven to be adequate in many applications by approximating nonlinear systems as linear systems in local regions of the $n$-dimensional state space
\cite{
slotine1991applied}.
We examine linear dynamical systems, as it is a necessary first step to understanding how target control may benefit nonlinear systems.
The linear time invariant (LTI) network dynamics are,
\begin{equation}\label{eq:system}
	\begin{aligned}
		\dot{\textbf{x}}(t) &= A \textbf{x}(t) + B \textbf{u}(t)\\[0.05in]
		\textbf{y}(t) &= C \textbf{x}(t)
	\end{aligned}
\end{equation}
where $\textbf{x}(t) = [x_1(t), \ldots, x_n(t)]^T$ is the $n \times 1$ time-varying state vector, $\textbf{u}(t) = [u_1(t), \ldots, u_m(t)]^T$ is the $m \times 1$ time-varying external control input vector, and $\textbf{y}(t) = [y_1(t), \ldots, y_p(t)]^T$ is the $p \times 1$ time-varying vector of outputs, or targets.
The $n \times n$ matrix $A = \{a_{ij}\}$ is the adjacency matrix described previously, the $n \times m$ matrix $B$ defines the nodes in which the $m$ control input signals are injected, and the $p \times n$ matrix $C$ expresses the relations between the states that are designated as the outputs.
In addition, the diagonal values of $A$, $a_{ii}$, $i=1,\ldots,n$, which represent self-regulation, such as birth/death rates in food webs, station keeping in vehicle consensus, degradation of cellular products, etc., are chosen to be unique at each node (see proposition 1 in \cite{cowan2012nodal}).
These diagonal values are chosen to also guarantee that $A$ is Hurwitz so the system in Eq. \eqref{eq:system} is internally stable.
We restrict ourselves to the case when $B$ ($C$) has linearly independent columns (rows) with a single nonzero element, i.e., each control signal is injected into a single node (defined as an input node) and each output is drawn from a single node (defined as a target node).
Our particular choice of the matrix $C$ is consistent with target control, as our goal is to individually control each one of the target nodes.
Our selection of the matrix $B$ is due to our assumption that different network nodes may be selectively affected by a particular control signal, e.g., a drug interacting with a specific node in a protein network.
Note that in today's information-rich world, a main technological limitation is not generating control signal, but rather placing actuators at the input nodes; hence our assumption, that each actuator is driven by an independent control signal is sound \cite{iudice2015structural}.
We define $\mathcal{P}_p \subseteq \mathcal{V}$ as the subset of target nodes and $p = |\mathcal{P}_p|$ as the number of target nodes.
A small sample schematic is shown in Fig. 1\textbf{a} that demonstrates the graphical layout of our problem emphasizing the graph structure and the role of input nodes and targets.
Here by an input node, we mean a node that directly receives one and only one control input such as nodes 1 and 2 in Fig. 1\textbf{a}.
%
% 1.c Solution to the LTI Continuous-time Problem
The explicit equation for the time evolution of the outputs is,
\begin{equation}\label{eq:output}
    \textbf{y}(t) = C e^{A(t - t_{\text{0}})} \textbf{x}_0 + C \int_{t_{\text{0}}}^{t} e^{A(t - \tau)} B \textbf{u}(\tau) d\tau,
\end{equation}
where we are free to choose $\textbf{u}(t)$ such that it satisfies the prescribed initial state, $\textbf{x}(t_{\text{0}}) = \textbf{x}_{\text{0}}$ and desired final output, $\textbf{y}(t_{\text{f}}) = \textbf{y}_{\text{f}}$.
Note that if we set $C = I_n$, where $I_n$ is the $n \times n$ identity matrix, then $\textbf{y}(t) = \textbf{x}(t)$.\\
%
%===========================================================
%% Section 2: Optimal Control Problems
% 2.a Minimum Energy Problem
\indent The minimum energy control input, well known from linear systems theory
\cite{
rugh1996linear}, 
minimizes the cost function $J = \frac{1}{2} \int_{t_{\text{0}}}^{t_{\text{f}}} \textbf{u}^T(t) \textbf{u}(t) dt$ and satisfies an arbitrary initial condition and an arbitrary final condition if the system is controllable.
A similar control input is optimal when the final condition is imposed on only some of the states, i.e., on the target nodes (see the derivation in Supplementary Note 2).
\begin{equation}\label{eq:umecs}
    \textbf{u}^*(t) = B^T e^{A^T(t_{\text{f}} - t)} C^T \left(C W C^T \right)^{-1} \left( \textbf{y}_{\text{f}} - e^{A(t_{\text{f}}-t_{\text{0}})} \textbf{x}_{\text{0}} \right)
\end{equation}
The real, symmetric, semi-positive definite matrix $W = \int_{t_{\text{0}}}^{t_{\text{f}}} e^{A(t_{\text{f}}-\tau)} B B^T e^{A^T(t_{\text{f}}-\tau)} d\tau$ is the controllability Gramian.
Note that in deriving Eq. \eqref{eq:umecs} we must assume that the triplet $(A,B,C)$ is output controllable, which can be determined if the matrix $\text{rank}(CB | CAB | \ldots | CA^{n-1}B) = p$.
If the triplet is output controllable, it implies that the matrix $CWC^T$ is invertible
\cite{
rugh1996linear,
murota1990note}.
This suggests the possibility that, while the entire network may not be controllable (i.e., $C = I_n$ and $W$ is singular), for a given $B$ (of the form described above) there may be a controllable subspace (subset of nodes) within the network.
On the other hand, every subspace of the controllable subspace is also controllable.
In the following discussions we proceed under the assumption that the pair $(A,B)$ is controllable by following the methodology in \cite{cowan2012nodal} and focus on the effect that the choice of the matrix $C$ has on the control energy.\\
%
% 2.b LQR Problem
\indent We also consider a more general linear-quadratic optimal control problem, i.e., we attempt to minimize a quadratic cost function that applies a weight to the states, $\textbf{x}(t)$, and the control inputs, $\textbf{u}(t)$.
This type of cost function is applied in a variety of science and engineering applications such as medical treatments or biological systems \cite{stengel2002optimal,chang2011inference}, consensus or synchronization of distributed agents \cite{cao2010optimal,cosby2012uncooperative,mosebach2014synchronization}, networked systems \cite{galvan2010hybrid}, social interactions \cite{bloembergen2014influencing}, and many more,
\begin{equation}\label{eq:cost}
    J = \frac{1}{2} \int_{t_{\text{0}}}^{t_{\text{f}}} \left[ \textbf{x}^T(t) Q \textbf{x}(t) + 2 \textbf{x}^T(t) M \textbf{u}(t) + \textbf{u}^T(t) R \textbf{u}(t) \right] dt.
\end{equation}
The $n \times n$ matrix $Q$ applies a weight to the states and the $m \times m$ matrix $R$ applies a weight to the control inputs.
The $n \times m$ matrix $M$ allows for mixed term weights which may arise for specially designed trajectories, optimization of human motion, or other physical constraints
\cite{bernstein2009matrix,
priess2015solutions,
chen2014inverse,
ali2015optimal}.
We restrict the cost function matrix $Q$ to be symmetric semi-positive definite and the matrix $R$ to be symmetric positive definite.
We derive a closed form expression for the optimal control input associated with Eq. \eqref{eq:cost} using the property that the Hamiltonian system which arises during the solution (derived in Supplementary Note 4) can be decoupled,
\begin{equation}\label{eq:umccs}
    \textbf{u}^*_{\text{c}}(t) = \underbrace{ -R^{-1} \left(M^T + B^T S \right) \textbf{x}(t)}_{\textbf{u}_{\text{c1}}^*(t) } + \underbrace{ -R^{-1} B^T e^{\tilde{A}^T(t_{\text{f}}-t)} C^T \left( C \tilde{W} C^T \right)^{-1} \left( \textbf{y}_f - C e^{\tilde{A}(t_{\text{f}}-t_{\text{0}})} \textbf{x}_{\text{0}} \right) }_{\textbf{u}_{\text{c2}}^*(t)}.
\end{equation}
The symmetric matrix $S$ is the solution to $S \bar{B} \bar{B}^T S - S \bar{A} - \bar{A}^T S - \bar{Q} = O_n$, the continuous time algebraic Riccati equation, and the other matrices are defined as,
\begin{equation}\label{eq:matrices}
    \begin{aligned}
        &\begin{aligned}
            \bar{A} = A - B R^{-1} M^T, && 
            \bar{B} = B \hat{R}^{-1/2}, &&
            \bar{Q} = Q - M R^{-1} M^T
        \end{aligned}\\[1mm]
        &\begin{aligned}
            \tilde{A} &= A - B R^{-1} M^T - B R^{-1} B^T S, &&
            \tilde{W} &= \int_{t_{\text{0}}}^{t_{\text{f}}} e^{\tilde{A} (t_{\text{f}} - \tau)} B \hat{R}^{-1} B^T e^{\tilde{A}^T(t_{\text{f}} - \tau)} d\tau
        \end{aligned}
    \end{aligned} 
\end{equation}
The derivation of Eqs. \eqref{eq:umccs} and \eqref{eq:matrices} is detailed in Supplementary Note 4.\\
%
%===========================================================
%% Section 3: Energy Discussion
% 3.a Definition
\paragraph{Optimal Energy and Worst Case Direction}
The energy associated with an arbitrary control input, such as Eq. \eqref{eq:umecs} or Eq. \eqref{eq:umccs}, while only targeting the nodes in $\mathcal{P}_p$, is defined as $E^{(p)} = \int_{t_{\text{0}}}^{t_{\text{f}}} \textbf{u}^T(t) \textbf{u}(t) dt$.
Note that $E^{(p)}$ also depends on which $p$ nodes are in the target set, $\mathcal{P}_p$, i.e., there is a distribution of values of $E^{(p)}$ for all target node sets of size $p$.
The energy $E^{(p)}$ is a measure of the `effort' which must be provided to achieve the control goal.
In the subsequent definitions and relations, when a variable is a function of $p$, we more specifically mean it is a function of a specific target set of size $p$ of which there are $\frac{n!}{p!(n-p)!}$ possible sets.
We can define the energy when the control input is of the form in Eq. \eqref{eq:umecs} as,
\begin{equation}\label{eq:emecs}
	E^{(p)} = \left( \textbf{y}_{\text{f}} - C e^{A(t_{\text{f}}-t_{\text{0}})} \textbf{x}_{\text{0}} \right)^T \left( C W C^T \right)^{-1} \left( \textbf{y}_{\text{f}} - C e^{A(t_{\text{f}}-t_{\text{0}}) \textbf{x}_{\text{0}}} \right) = \boldsymbol{\beta}^T W_p^{-1} \boldsymbol{\beta}
\end{equation}
where the vector $\boldsymbol{\beta} = \textbf{y}_{\text{f}} - C e^{A(t_{\text{f}}-t_{\text{0}})} \textbf{x}_{\text{0}}$ is the \emph{control maneuver} and $W_p$ is the $p \times p$ symmetric, real, non-negative definite output controllability Gramian.
Note that when $C$ is defined as above, i.e., its rows are linearly independent versors, the \emph{reduced Gramian} $W_p$ is a $p$-dimensional principal submatrix of $W$.
%
% 3.b Figure 1 Example Discussion
A small, three node example of the benefits of target control is shown in Fig. 1(\textbf{b})-(\textbf{g}).
In the first scenario, Fig. 1(\textbf{b})-(\textbf{d}), each node has a prescribed final state ($p = n = 3$) and in the second scenario 1(\textbf{e})-(\textbf{g}) only a single node is targeted ($p = 1$).
The energy is calculated for each scenario by integrating the curves in Figs. 1(\textbf{d}) and 1(\textbf{g}) from which we find that $E^{(3)} = 382$ and $E^{(1)} = 66.3$.
Even though the second scenario has one third of the targets, the energy is reduced by a sixth (compare also the different scales on the y-axis of Figs. 1(\textbf{d}) and 1(\textbf{g})).
%
% 3.c Maximum Energy
We denote the eigenvalues of $W_p$ as $\mu_i^{(p)}$, $i = 1, \ldots, p$, which are ordered such that $0 < \mu^{(p)}_1 \leq \ldots \leq \mu^{(p)}_p$ when the triplet ($A,B,C$) is output controllable.
By defining the magnitude of the vector, $|\boldsymbol{\beta}| = \beta$, we can define the `worst-case' (or maximum) energy according to the Min-Max theorem which provides a bound for Eq. \eqref{eq:emecs}.
The bounds are functions of the extremal eigenvalues of $W_p$,
\begin{equation}\label{eq:minmax}
	0 < \frac{\beta^2}{\mu^{(p)}_p} \leq \boldsymbol{\beta}^T W_p^{-1} \boldsymbol{\beta} \leq \frac{\beta^2}{\mu^{(p)}_1} < \infty.
\end{equation}
The upper extreme of the control energy for any control action is $\max \left\{ E^{(p)} \right\} \sim \frac{1}{\mu^{(p)}_1}$, which is what we call the `worst-case' energy.
%
% 3.d Average Energy
For an arbitrary vector $\boldsymbol{\beta}$, which can be represented as a linear combination of the eigenvectors of $W_p$, the energy can be defined as a weighted sum of the inverse eigenvalues, $1/\mu^{(p)}_i$, which includes the worst-case energy.
Moreover, for the large scale-free networks that are of interest in applications, typically $\mu^{(p)}_1 << \mu^{(p)}_j$, $j = 2, \ldots, p$, and $1/\mu_1^{(p)}$ provides the approximate order of the energy required to move the system in any arbitrary direction of state space.
This is demonstrated with an example in Supplementary Note 5.\\
%
%===========================================================
%% 4) Energy Scaling Derivation
% 4.a Eigenvalue Relationships
\indent We investigate how the selection of the target nodes affects $E_{\max}^{(p)}$, the inverse of the smallest eigenvalue of the output Gramian.
In order to better understand the role of the number of target nodes on the worst-case energy, we consider an iterative process by which we start from the case when every node is in the target set, $\mathcal{P}_n = \mathcal{V}$, and progressively remove nodes.
Say $\mu_j^{(i)}$ ($\mu_j^{(i-1)}$) is an eigenvalue of $W_{i}$ before (after) removal of a target node.
By Cauchy's interlacing theorem we have that,

\begin{equation}\label{eq:CIT}
    0 < \mu_1^{(i)} \leq \mu_1^{(i-1)} \leq \mu_2^{(i)} \leq \mu_2^{(i-1)} \leq \ldots \leq \mu_{i-1}^{(i)} \leq \mu_{i-1}^{(i-1)} \leq\mu_i^{(i)}
\end{equation}
In particular, from \eqref{eq:CIT}, we note that $\mu_1^{(i)} \leq \mu_1^{(i-1)}$, indicating that the smallest eigenvalue cannot decrease after removal of a target node.
This implies that the maximum energy $E_{\max}^{(i)} \geq E_{\max}^{(i-1)}$ for all $i$ such that $ 1 \leq i \leq n-1$.  
%
% 4.b Scaling Iterative Process
\paragraph{Energy Scaling with Reduction of Target Space}
We would like to determine the rate of increase of $\mu_1^{(p)}$ as $p$ decreases which is not obvious from Eq. \eqref{eq:CIT}.
At each step $p$, $\mathcal{P}_p$ contains $p$ nodes in the target set (such that $\mathcal{P}_{p} \subset \mathcal{P}_{p+1}$ and $p$ decreases from $n-1$ to $1$) and the output controllability Gramian is partitioned such that $W_p$ is a principal minor of $W_{p+1}$.
\begin{equation}\label{eq:Witer}
    W_{p+1} = \left[ \begin{array}{cc}
        w_{pp} & \textbf{w}_{p}^T\\
        \textbf{w}_{p} & W_p
    \end{array} \right]
\end{equation}
We let the matrix $\bar{W}_p$ be the matrix $W_{p+1}$ except that the first row of $W_{p+1}$ in Eq. \eqref{eq:Witer} has been replaced with zeros, and we define the vectors $\textbf{v}_p$ ($\bar{\textbf{v}}_p$) to be the left (right) eigenvector associated with the smallest eigenvalue of $W_p$ ($\bar{W}_p$).
The relation between two consecutive values, $\mu_1^{(p)}$ and $\mu_1^{(p+1)}$, can be expressed linearly as $\mu_1^{(p)} = \mu_1^{(p+1)} \eta_p$ where $\eta_p = 1 - \frac{\left[ \textbf{v}_{p} \right]_1 \left[ \bar{\textbf{v}}_{p+1} \right]_1}{\textbf{v}_p^T \bar{\textbf{v}}_{p+1}} \geq 1$.
The notation $\left[ \textbf{a} \right]_1$ denotes the first value of a vector $\textbf{a}$.
Each value of $\eta_p$ exactly quantifies the rate of increase at each step of the specific process and also relates the maximum energies $E_{\max}^{(p+1)} = E_{\max}^{(p)} \eta_p$.
%
% 4.c Developing the Equation for Eta
We can also relate any two target sets of size $k$ and $j$ such that $1 \leq k < j \leq n$ and $\mathcal{P}_k \subset \mathcal{P}_j$,
\begin{equation}\label{eq:bareta}
    \log E_{\max}^{(j)} - \log E_{\max}^{(k)} = \sum_{i=k}^{j-1} \log \eta_i = (j-k) \log \bar{\eta}_{(k \rightarrow j)}
\end{equation}
where $\bar{\eta}_{(k \rightarrow j)}$ is the geometric mean of $\eta_i$, $i = k, \ldots, (j-1)$, which is \emph{independent} of the order of the nodes chosen to be removed between $\mathcal{P}_k$ and $\mathcal{P}_j$.
To define a network characteristic parameter $\eta$, we average Eq. \eqref{eq:bareta} over many possible choices of the target sets $\mathcal{P}_k$ and $\mathcal{P}_j$, where we have selected $k = n/10$ and $j = n$,
\begin{equation}\label{eq:eta}
    \eta \equiv n \left\langle \log \bar{\eta}_{(\frac{n}{10} \rightarrow n)} \right\rangle
\end{equation}
where the symbol $\left\langle \cdot \right\rangle$ indicates an average over many possible choices of $n/10$ nodes for the target set.
By applying Eq. \eqref{eq:eta} to Eq. \eqref{eq:bareta} and by setting $k = n/10$ and $j = p > k$ (for the an extended discussion see Supplementary Note 3), we achieve the scaling equation used throughout the simulations,
\begin{equation}\label{eq:energy_scale}
    \left\langle \log E_{\max}^{(p)} \right\rangle \sim \frac{p}{n} \eta.
\end{equation}
The linear relationship is shown in Figs. 2, 3, and 4, where $\frac{p}{n}$ is decreased from 1 (the target set $\mathcal{P}_n = \mathcal{V}$) to 0.1 (the target set consists of $10\%$ of the nodes drawn randomly from the set of all nodes).
Further details of the scaling law and its relation to the spectral characteristics of the output controllability Gramian can be found in Supplementary Note 3 and the practical calculation can be found in the Methods.
For the simulations in Figs. 2,3,4,5,6 and 7, around 50\% of the nodes are chosen to be input nodes (which we have verified yields a controllable pair $(A,B)$).\\
%
%===========================================================
%% 5) Model Network Analysis
% 5.a Introduction
\indent The exponential decay of the energy as $p/n$ decreases has immediate practical relevance as it indicates that large networks which may require a very large amount of energy to fully control \cite{yan2015spectrum}, will require much less for even significant \emph{portions} of the network.
However, the rate of this exponential decrease, $\eta$, is network specific.
We compute the value of $\eta$ for fifty scale-free model networks, constructed with the static model in Ref. \cite{goh2001universal} for specific parameters $k_{\text{av}}$, the average degree, and $\gamma_{\text{in}} = \gamma_{\text{out}} = \gamma$, the power law exponent of the in- and out-degrees, and take the mean over the realizations.
%
% 5.b Figure 2 Discussion
We see in Fig. 2 that $\eta$ varies with both of the network parameters $\gamma$ and $k_{\text{av}}$.
A large value of $\eta$ indicates that target control is highly beneficial for that particular network, i.e., the average energy required to control a portion of that network is much lower when the size of the target set is reduced.
In Figs. 2(\textbf{a}) and 2(\textbf{b}), the exponentially increasing value of the worst-case energy $E_{\max}^{(p)}$ is shown with respect to the size of the target set normalized by the size of the network, $p/n$, for various values of $\gamma_{\text{in}} = \gamma_{\text{out}} = \gamma$ when $k_{\text{av}} = 2.5$ and $8.0$, respectively.
The bars in Figs. 2(\textbf{a}) and 2(\textbf{b}) are one standard deviation over the fifty realizations each point represents, or in other words, when $p$ nodes are in the target set $\mathcal{P}_p$, it is most likely that $E_{\max}^{(p)}$ will lie between those bars.
The decrease of $\eta$ as $\gamma$ and $k_{\text{av}}$ increase for scale-free networks is displayed in Fig. 2(\textbf{c}).
Overall, we see that $\eta$ is largest for sparse, nonhomogeneous networks (i.e., low $k_{\text{av}}$ and low $\gamma$) which are also the `hardest' to control, i.e., they have the largest worst-case energy when all of the nodes are targeted.
This indicates that target control will be particularly beneficial when applied to metabolic interaction networks and protein structures, some of which are symmetric and which are known to have low values of $\gamma$ \cite{albert2002statistical}, as seen in Fig. 2(\textbf{b}), where both classes of networks are shown to have large values of $\eta$.\\
%
% 5.c Figure 3 Discussion
\indent The effects other network parameters have on $\eta$ are examined in Fig. 3.
Figure 3(\textbf{a}) displays some sample curves for $E_{\max}^{(p)}$ for shorter or longer values of $(t_{\text{f}} - t_{\text{0}})$, the time horizon.
The inset shows how $\eta$ increases as the time horizon $(t_{\text{f}} - t_{\text{0}})$ decreases.
We see that when $(t_{\text{f}}-t_{\text{0}})$ approaches zero from the right, $\eta$ increases sharply, which shows the increased benefit of target control as the time horizon is reduced.
Figure 3(\textbf{b}) examines how $E_{\max}^{(p)}$ changes for various numbers of input nodes (represented as a fraction of the total number of nodes in the network).
The inset collects values of $\eta$ for different values of $n_{\text{d}}$, which increases as the number of input nodes is decreased.
The role of the time horizon \cite{yan2012controlling} and the number of input nodes \cite{yan2015spectrum} on the control energy have been discussed in the literature for the case in which all the nodes were targeted.\\
\indent Comparing the results between both panels in Fig. 3 and the results in Fig. 2, we see that each parameter has more or less of an effect on the control energy.
Shortening the time horizon from the nominal value $t_{\text{f}} = 1$ (which was used in Fig. 2) by four orders of magnitude doubled the value of $\eta$.
Decreasing the value number of input nodes from $n/2$ (the number used in Fig. 2) to only $n/5$ also roughly doubled the value of $\eta$.
In comparison, increasing the heterogeneity of the network, by decreasing the power-law exponent $\gamma$, from three to slightly larger than two increased $\eta$ ten to twenty fold.
Clearly the underlying topology, as described by the power-law exponent, plays the largest role in determining (and thus affecting) the control energy.\\
%
%% 6) Real Dataset Analysis
% 6.a Introduction
\indent We also analyze datasets collected from various fields in science and engineering to study how the worst-case energy changes with the size of the target set for networks with more realistic structures.
We are particularly interested in the possibility that these networks display different properties in terms of their target controllability, when compared to the model networks analyzed.
To this end, we consider different classes of networks, e.g., food webs, infrastructure, metabolic networks, social interactions, etc.
The name, source, and some important properties of each of the datasets are collected in Supplementary Note 8.
For each network we choose edge weights and diagonal values from the uniform distribution as discussed in the Methods section below.
%
% 6.b Figure 4 Discussion
Overall we see a similar relationship in terms of the average degree $k_{\text{av}}$ and $\eta$ in Fig. 4(\textbf{c}) as for the model networks in Fig. 4(\textbf{c}).
The real datasets which have a large worst-case energy when all of the nodes are targeted, $E_{\max}^{(n)}$, tend to also have the largest value of $\eta$ which acts as a measure of the rate of improvement with target control.
It should be noted that the value of $\eta$ varies little within each class of networks (e.g., food webs, infrastructure, metabolic networks, social interactions, etc. as seen in Fig. 4(\textbf{c})) which suggests that the structure of each class is similar.
Fields of study where networks tend to have a large $\eta$ would benefit the most from examining situations when a control law could be implemented that only targets some of the elements in the network.\\
\indent For an arbitrary network, $\eta$ cannot be accurately determined from a single value of $E_{\max}^{(p)}$ as some networks which have a large worst-case energy when every node is targeted can have a much smaller worst-case energy when only a small portion of the network is controlled as compared to other networks.
It is interesting to note from Figs. 4(\textbf{a}) and 4(\textbf{b}) that at some target fraction $p/n$ the energy trends of two different real networks may cross.
Specifically, in Fig. 4(\textbf{a}), when every node is targeted, $p/n = 1$, the \textit{s420st} \cite{milo2004superfamilies} circuit has a larger maximum energy, $E^{(n)}_{\max}$, than the \textit{TM-met} \cite{jeong2000large} metabolic network.
However, when $p/n$ is smaller than $0.6$, it requires, on average, more energy to control a portion of the \textit{TM-met} network than an equivalent portion in the \textit{s420st} network.
The same type of behavior is seen in Fig. 4(\textbf{b}) between three networks: Food web \textit{Carpinteria} \cite{lafferty2006food}, a protein interaction network \textit{prot\_struct\_1} \cite{milo2004superfamilies} and social network \textit{FB forum} \cite{opsahl2013triadic}.
In summary, we can see that one can estimate the value of $\eta$ from the average degree of the network but to determine the worst-case energy, at least one point along the energy curve for a specific cardinality of the target set is also required (as in Figs. 4(\textbf{a}) and 4(\textbf{b})).\\
%
% 6.c Statistical Analysis (Figure 5)
\indent Figure 5 shows a comparison for several real networks between the value of $\eta$ of each original network and the values of $\eta$ for an ensemble of networks that have been generated by randomly rewiring each real network's connectivity while preserving the degrees of its nodes (see Methods).
We see that for all the real networks examined, their value of $\eta$ is larger than the values of $\eta$ obtained for the randomized versions to a statistically significant level.
We conclude that the potential advantage of applying target control to real networks is higher than for networks derived from random connections such as the static model which we have used to construct our model networks.\\
%
%% 7) LQ Energy Formulation
% 7.a Introduction
\indent We compute the energy for the control input $\textbf{u}_{\text{c}}^*(t)$.
The control consists of two parts, $\textbf{u}_{\text{c1}}^*(t)$ which is proportional to the states and $\textbf{u}_{\text{c2}}^*(t)$ which is of a similar form to Eq. \eqref{eq:umecs}.
\begin{equation} \label{eq:emccs}
    \begin{aligned}
	    E_{\text{c}}^{(p)} &= \int_{t_{\text{0}}}^{t_{\text{f}}} \left( \textbf{u}^*_{\text{c1}}(t) + \textbf{u}^*_{\text{c2}}(t) \right)^T \left( \textbf{u}^*_{\text{c1}}(t) + \textbf{u}^*_{\text{c2}}(t) \right) dt\\[0.1in]
	    &= \int_{t_{\text{0}}}^{t_{\text{f}}} \left[ \textbf{u}^{*T}_{\text{c1}}(t) \textbf{u}^*_{\text{c1}}(t) + 2 \textbf{u}^{*T}_{\text{c1}}(t) \textbf{u}^*_{\text{c2}}(t) \right] dt + \int_{t_{\text{0}}}^{t_{\text{f}}} \textbf{u}_{\text{c2}}^{*T}(t)\textbf{u}^*_{\text{c2}}(t) dt
    \end{aligned}
\end{equation}
Note that the second integral in the second line of Eq. \eqref{eq:emccs}, when $R = I_m$, is the quadratic form $\tilde{\boldsymbol{\beta}}^T \tilde{W}_p^{-1} \tilde{\boldsymbol{\beta}}$ which scales exponentially with the cardinality of the target set.
The other two terms are functions of the state trajectory which are not appreciably altered by the number of targeted nodes.
We thus expect to see similar energy scaling behavior for the cost function Eq. \eqref{eq:cost} with $Q\neq O_{n \times n}$ and $M \neq O_{n\times m}$.\\
%
% 7.b Figure 6 Discussion
\indent In some applications a cost applied to the states may be beneficial as it will substantially alter the state trajectories (see the example in Supplementary Note 4).
In the following simulations, to restrict the number of variables we consider, the mixed term weight matrix $M = O_{n\times m}$ and the state weight matrix $Q = \zeta I$, i.e., a diagonal matrix with constant real value, $\zeta$, on the diagonal.
In Fig. 6(\textbf{a}) model networks are considered of different scale-free exponents $\gamma$.
In 6(\textbf{b}), the real networks \textit{IEEE 118 bus test grid} \cite{IEEE} and \textit{Florida everglades foodweb} \cite{pajek} are optimally controlled with respect to the cost function in Eq. \eqref{eq:cost}, and the approximate maximum energy (computed by numerically integrating Eq. \eqref{eq:cost}) is determined for increasing values of the scalar $\zeta$.
As $\zeta$ increases in Figs. 6(\textbf{a}) and 6(\textbf{b}), each point along the curve is of approximately the same order of magnitude.
As $\zeta$ is varied, the order of magnitude of the maximum energy does not change substantially, and mainly depends on the triplet $(A,B,C)$ without much effect by the matrix $Q$.\\
%
% 7.c Figure 7 Discussion
Finally, we offer evidence to connect the energy scaling law derived for the minimum energy optimal control problem to the energy scaling apparent for the control signal that arises in the solution of the general quadratic cost function, Eq. \eqref{eq:cost}.
Figure 7 shows that not only does the order of magnitude of the maximum energy not change significantly, but the rate of increase, $\eta$, of the maximum energy does not change significantly with respect to the size of the target set either.
We compute $\eta$, the energy scaling, for a single model network while we increase the state weight cost matrix defined as the diagonal matrix $Q = \zeta I_n$.  
This suggests that if $\eta$ is computed for a network with respect to the minimum energy formulation, it can be used to approximate $\eta$ when the cost function is quadratic with respect to the states as well.
%
%% End of Results Section
%
%% Begin Discussion Section
\section*{Discussion}
\indent This paper discusses a framework to optimally control a portion of a complex network for assigned initial conditions and final conditions, and given the sets of input nodes and target nodes.
We provide an analytic solution to this problem in terms of a reduced Gramian matrix $W_p$, where the dimensions of this matrix are equal to the number of target nodes one attempts to control.
We show that for a fixed number of input nodes, the energy required to control a portion of the network decreases exponentially when the cardinality of the target so even controlling a significant number of nodes requires much less energy than when every node is targeted.
The energy reduction, expressed as the rate $\eta$, is largest for networks which are heterogeneous (small power-law exponent $\gamma$ in a scale-free degree distribution) and sparse (small $k_{\text{av}}$), with a short time horizon and fewer control inputs.
The control of these networks typically has especially large control energy demands.
Thus target control is most beneficial for those networks which are most difficult to control.
From the simulations that we have performed on model networks, we have seen that the effect each of these parameters has is not equal.
The control energy required is most dependent on the underlying structure of the network which we see can increase $\eta$ by as much as twenty times holding all other parameters constant.
Adjusting the time horizon over multiple orders of magnitude, or reducing the number of input nodes from 50\% to 20\% doubled the value of $\eta$, which is a comparatively small increase.\\
\indent The potential applications for developing target controls are numerous, from local jobs among networked robots to economic policies designed to affect only specific sectors.
We see that datasets from the literature in many fields also experience the reduced energy benefits from target control.
The networks which describe metabolic interactions and protein structures have some of the largest values of $\eta$ suggesting target control would by the most beneficial in those fields.\\
\indent We have also considered a linear-quadratic optimal control problem (in terms of the objective function \eqref{eq:cost}) applied to dynamical complex networks.
We show that the scaling factor $\eta$ for a network with control parameters $n_{\text{d}}$ and $t_{\text{f}}$ remains nearly the same whether the control is optimal with respect to the minimum energy control input as in Eq. \eqref{eq:umecs} or is optimal with respect to the quadratic cost function in Eq. \eqref{eq:cost} as in Eq. \eqref{eq:umccs}.
The observed decrease of the control energy over many orders of magnitude indicates a strong potential impact of this research in applications where control over the entire network is not necessarily required.
%
%% End of Discussion Section
%
%% Begin Methods
%
%% 1) Model Network Construction
%
\subsection{Model Networks.}
In our analyses, similar to \cite{yan2015spectrum}, we assume the networks have stable dynamics.
The scale free model networks we consider throughout the paper and the supplementary information are constructed with the static model \cite{goh2001universal}.
The Erdos-Renyi graphs represent the static model when the nodal weights are all the same, i.e., when the power-law exponent approaches infinity.
Edge weights are chosen from a uniform distribution between $0.5$ and $1.5$.
Diagonal noise, $\delta_i$, is included, drawn from a uniform distribution between $-1$ and $1$ so that the eigenvalues of the adjacency matrix are all unique.
The weighted adjacency matrix $A$ is stabilized with a value $\epsilon$ such that each diagonal value of $A$ is $\{a_{ii}\} = \delta_i + \epsilon$ where $i = 1, \ldots, n$.
The value $\epsilon$ is chosen such that the maximum eigenvalue of $A$ is equal to $-1$.
The matrix B is constructed by choosing which nodes in the network require an independent control signal.
The unique diagonal values of the adjacency matrix ensure that only source nodes (those with no incoming connections), and one node from each strongly connected component, require these control signals \cite{cowan2012nodal}.
These nodes are used to create the set of driver nodes, i.e., those which received a control signal directly (see Fig. 1(a)).
Additional nodes are added to the set of driver nodes randomly until the desired number of driver nodes is reached.
The matrices $B$ ($C$) are composed of $m$ ($p$) versors as columns (rows).
The controllability Gramian, $W_p$, can be calculated as a function of the eigendecomposition of the state matrix $A = V \Lambda V^{-1}$,
\begin{equation}\label{eq:gram_calc}
    W_p = C V \left( Y \circ V^{-1} B B^T V^{-T} \right) V^T C^T
\end{equation}
where the notation $V^{-T}$ denotes the transpose of the inverse of a matrix $V$.
Note that $V$ must be invertible (so that $A$ is diagonalizable), i.e., the eigenvectors of $A$ must span $\mathbb{R}^n$.
The matrix $Y$ has elements,
\begin{equation}\label{eq:int_matrix}
    Y_{ij} = \frac{\exp \left[ (\lambda_i + \lambda_j) (t_f - t_0) \right] - 1}{\lambda_i + \lambda_j}
\end{equation}
Note that the uniqueness and negative definiteness of the eigenvalues ensures that $Y_{ij}$ is finite for every $i,j = 1, \ldots, n$, i.e., $\lambda_i + \lambda_j \neq 0$, and the set of eigenvectors of $A$ are linearly independent and thus the inverse of $V$ exists.\\
%
%% 2) Choosing Driver Nodes
%
\subsection{Choosing Input Nodes.}
When determining the set of input nodes that guarantees network controllability, often the methods presented in Ref. \cite{liu2011controllability}, derived from structural controllability, are applied.
As the networks we are concerned with have unique diagonal elements in the adjacency matrix, structural controllability states that the network can be controlled with a single control input attached to every node in the network (see theorem 1 and proof in \cite{cowan2012nodal}).
Ref. \cite{cowan2012nodal} considers an adjacency matrix with unique diagonal elements along the main diagonal and states that this type of matrix can be controlled with a single control input attached to the power-dominating set (PDS) of the underlying graph.
The PDS is the smallest set of nodes from which all other nodes can be reached, i.e., there is at least one directed path from the nodes in the PDS to every other node in the network.
In the work presented here, different from \cite{cowan2012nodal}, we compute an over-estimate of the PDS (that retains the property that all other nodes in the network are reachable) and attach a \emph{unique control input} to each node in the set.
We then add additional nodes, chosen randomly, to the set of input nodes until there are $m$ input nodes where $m$ is pre-defined integer less than $n$.
Thus, if there are $m$ input nodes, then there are $m$ control inputs (see the sample network in Fig. 1(\textbf{a})).
%
%% 3) Determining \eta numerically
%
\subsection{Practical Computation of $\eta$.}
\indent Here we provide additional details on how Figs. 2, 3, and 4, which show the exponential scaling of the energy with respect to the cardinality of the target set, were generated.
For large networks, computing the mean over all possible sets of target nodes is computationally expensive.
Instead, we approximate $\eta$ by computing the mean value of $\log E_{\max}^{(p)}$ for some sample values of $p$, $p = n/10, 2n/10, \ldots, n$ by randomly choosing $p$ nodes to be in a target set and computing the inverse of the smallest eigenvalue of $W_p$.
In each of the simulations, we compute the mean and standard deviation of the logarithm of the smallest eigenvalue of $W_p$ for typically 50 iterations.
By plotting the values of $\left\langle \log E_{\max}^{(p)} \right\rangle$, we see that a linear model is appropriate and we compute a linear least-squares best fit for the data.
The linear curve fit provides a good approximation of $\log E_{\max}^{(p)}$ as shown in Figs. 2, 3, and 4.
%
%
%% 4) Degree Preserving Randomization (DPR)
%
\subsection{Degree Preserving Randomization.}
To test whether the value of $\eta$ measured for the real networks is a function of just the average degree, $k_{\text{av}}$, and degree distribution (scale-free, exponential, etc.) or if there are other factors which play a role, we measure $\eta$ for randomized versions of the real networks.
We use degree preserving randomization (DPR) to ensure that the randomized real network has the same average degree and the same degree sequence.
The randomization `rewires' the edges of the network by randomly choosing two edges and swapping the receiving node.
The process is repeated for an allotted amount of iterations until the networks are sufficiently rewired.
We compare each real network with its rewired counterparts in terms of their measured values of $\eta$.
We see in every case that $\eta_{\text{real}}$, the value of $\eta$ which corresponds to an original network derived from a dataset listed in Supplementary Table 1, deviates significantly from the distribution of $\eta$ for the DPR networks.
The corresponding p-values are listed in Fig. 5.
The disparity indicates that the real networks have special network features unaccounted for in the randomly rewired versions.
Furthermore, because for all cases $\eta_{\text{real}}$ is greater than any $\eta$ obtained from the DPR networks, our target strategies are more beneficial for the original networks.
%
% 5) Issues with Numerical Procedures
%
\subsection{Numerical Controllability.}
Recent literature on the control of complex networks has discussed the importance of recognizing the differences between theoretically controllable networks and numerically controllable networks.
The issue arises in Gramian based control schemes as the condition number of the Gramian can be quite large for certain `barely' controllable systems, i.e., ones where the control inputs only just satisfy analytic controllability measures.
Ref. \cite{
sun2013controllability}
found a second phase transition after a system ($A$,$B$) becomes analytically controllable, named the \emph{numerical controllability transition}.
While we acknowledge the importance of recognizing the second transition, for this article, we opt to use the multi-precision package Advanpix for Matlab so we can examine trends even when there is a relatively small number of control inputs which would otherwise make some networks be not numerically controllable using double precision.
For example, the Matlab toolbox Advanpix \cite{advanpix} allows the computation of the eigendecomposition of the Gramian $W$ to be performed in an arbitrarily precise manner.
Say $\mu_i$ and $\boldsymbol{v}_i$ are the $i$th eigenvalue and eigenvector, respectively.
The average residual error, using Advanpix, is,
\begin{equation}
    \left\langle \left| W \boldsymbol{v}_i - \mu_i \boldsymbol{v}_i \right| \right\rangle = \mathcal{O}(10^{-a})
\end{equation}
Typical values of $a$ used throughout this paper are 100 to 200.\\
We also use Advanpix when computing the energy for the general quadratic cost function in Eq. \eqref{eq:emccs}.
To approximate the integral, we use Legendre-Gauss (LG) quadrature with appropriate weights and points.
\begin{equation}
    E_c^{(p)} = \int_{t_0}^{t_f} \boldsymbol{u}^{*T}_{c}(t)\boldsymbol{u}^*_c(t) dt \approx \frac{t_f-t_0}{2} \sum_{i=1}^L w_i \boldsymbol{u}^{*T}_c(\tau_i) \boldsymbol{u}^*_c(\tau_i)
\end{equation}
We choose $L=50$ and compute the necessary LG weights $w_i$ and LG points $\tau_i$, $i=1,\ldots,50$.
%
% 6) Data Availability
%
\subsection{Data availability.}
The codes used to obtain the results in this study are available from the authors on reasonable request.
%
%% End of Methods Section
%
%% Begin of End Notes
\begin{addendum}
\item 
We gratefully acknowledge support from the National Science Foundation through NSF grant CMMI- 1400193, NSF grant CRISP- 1541148 and from the Office of Naval Research through award No. N00014-16-1-2637.
We thank Franco Garofalo, Francesco Lo Iudice, Jorge Orozco, Elvia Beltran Ruiz, Jens Lorenz, and Andrea L'Afflitto for insightful conversations.
\item[Author Contributions]
A.S., I.K., and F.S. formulated the problem statement.
A.S. and I.K. performed the mathematical analysis and numerical simulaions.
A.S., I.K., and F.S. wrote the paper.
F.S. supervised the research.
\item[Competing Interests]
The authors state there is no conflict of interest.
\item[Correspondence]
Reprints and permissions information is available at www.nature.com/reprints. The authors declare no competing financial interests. Correspondence and requests for materials should be addressed to I.K. (iklick@unm.edu) or F.S. (fsorrent@unm.edu). 
\item[Supplementary Information] is available in the online version of the paper.
\end{addendum}
%
%% End of End Notes
%
%% Begin of Bibliography

%
%% End of Bibliography
%
%% Begin of Figure Captions
%% Figure 1
\begin{figure}
\centering
\includegraphics[width=0.5\textwidth]{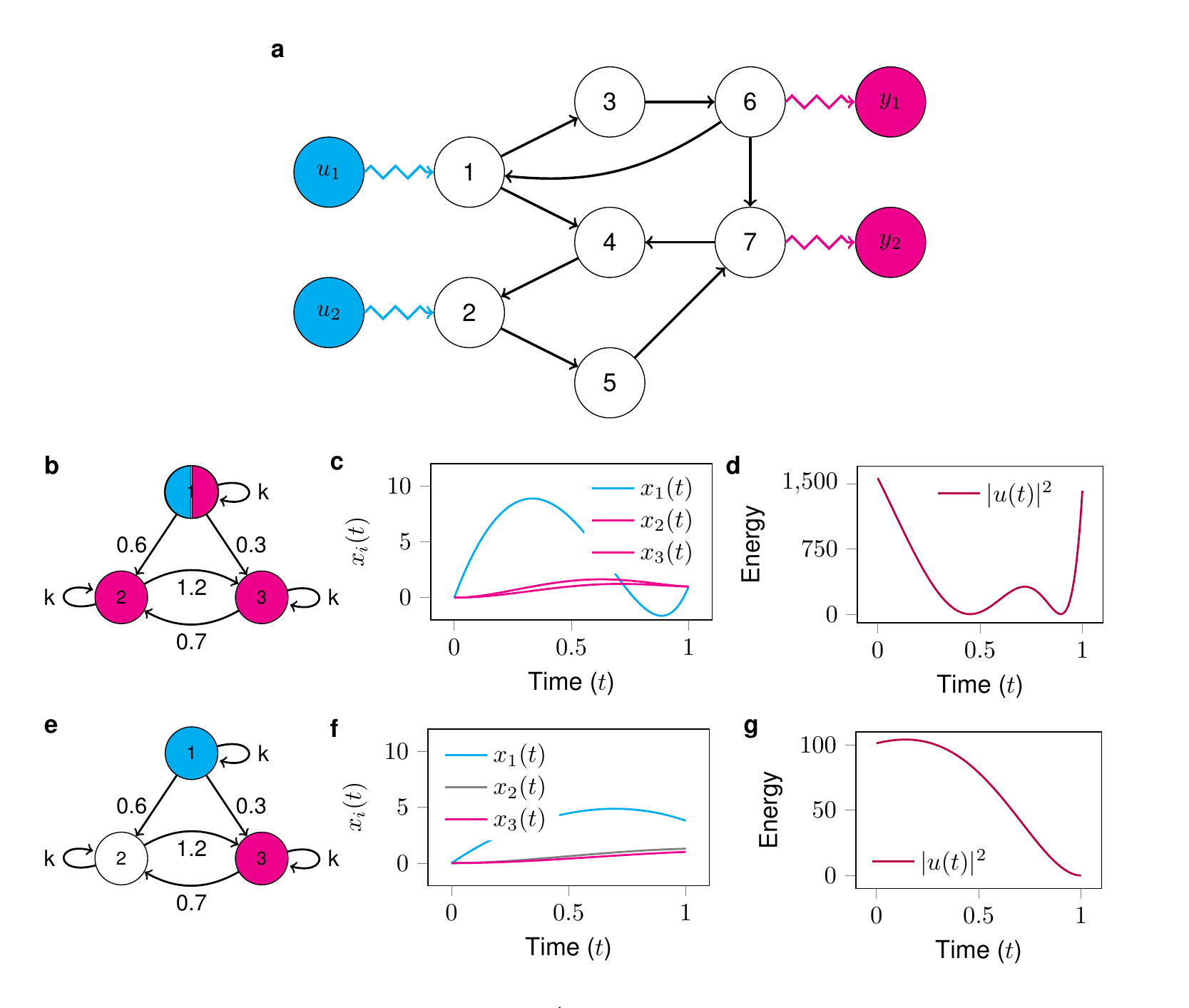}
\caption{
\textbf{An example network and control energy reduction with fewer targets.}
(\textbf{a}) A sample network with seven nodes and color-coded input signals (blue) and output sensors (pink).
Note that each control input is directly connected to a \emph{single} node, and each output sensor receives the state of a \emph{single} node.
Nodes directly connected to the pink outputs are \emph{target nodes}, i.e., they have a prescribed final state that we wish to achieve in finite time, $t_{\text{f}}$.
The corresponding vector ${\textbf{y}}(t_{\text{f}})$ is defined in terms of the states as well.
Nodes directly receiving a signal from a blue node are called \emph{input nodes} and the remaining nodes are neither input nodes nor target nodes.
(\textbf{b}) We examine a three node network where every node is a target node (pink nodes) and one node receives a control input (blue).
The edge weights are shown and the self-loop magnitude $k=1$.
(\textbf{c}) The state evolution is shown where the initial condition is the origin and the final state for each target node is $y_i(t_{\text{f}}) = 1$, $i=1,2,3$.
(\textbf{d}) The square of the magnitude of the control input is also shown.
The energy, or the control effort, is found by integrating the square of the magnitude of the the control input.
For this case, $E = \int |u(t)|^2 \approx 382$ (a.u.).
(\textbf{e}) The same network as in (\textbf{b}) but now only one node is declared a target node.
(\textbf{f}) The state evolution is shown where the initial condition remains the origin but the final condition is only defined for $y_3(t_{\text{f}}) =1 $.
(\textbf{g}) The square of the magnitude of the control input is also shown.
Note the different vertical axis scale as compared to (\textbf{d}).
For the second case, $E = \int |u(t)|^2 \approx 66.3$ (a.u.).}
\end{figure}
%
%% Figure 2
\begin{figure}
\centering
\includegraphics[width=0.5\textwidth]{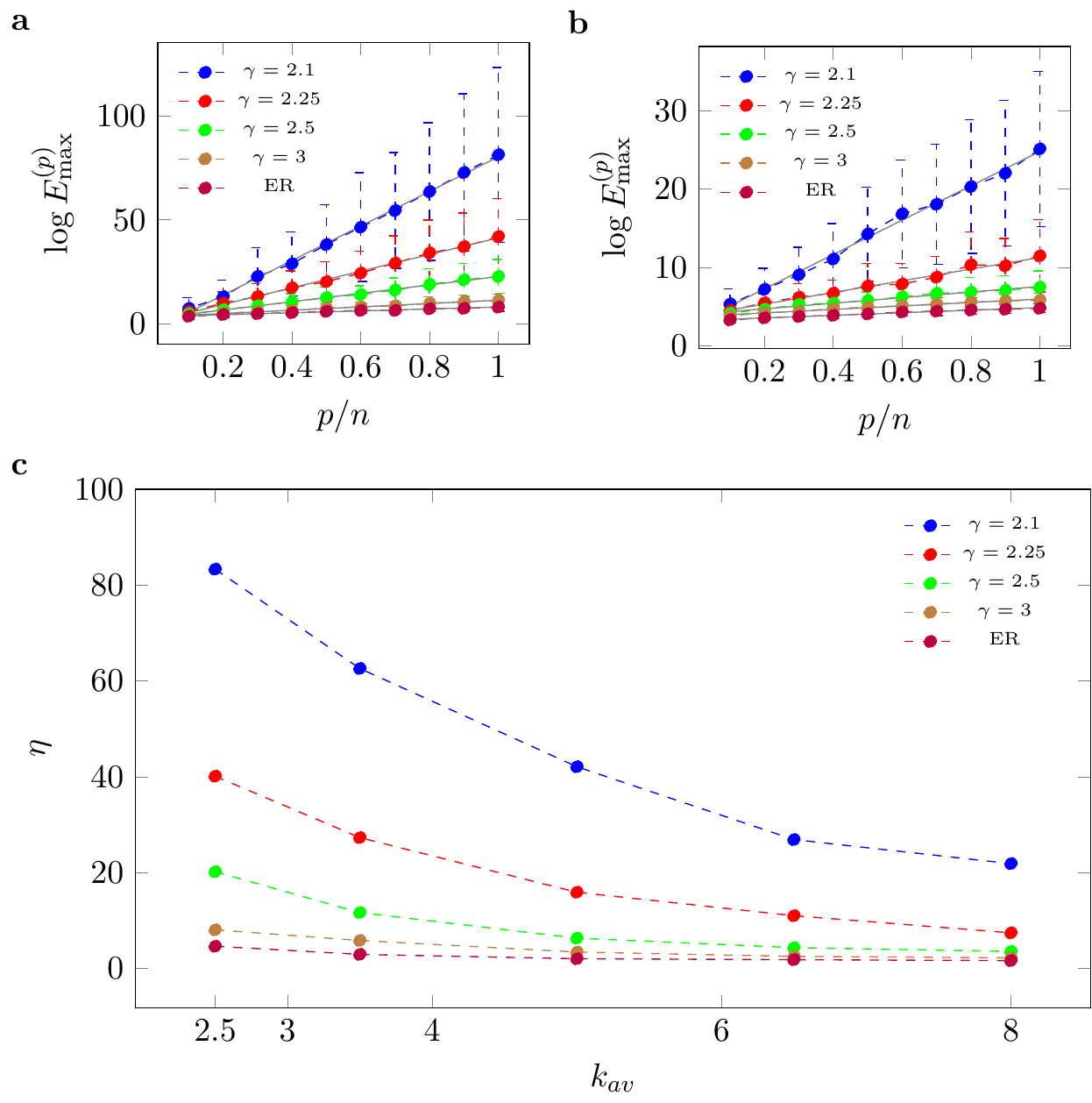}
\caption{
\textbf{The variation of $\eta$ with respect to model network parameters.}
(\textbf{a}) The maximum control energy is computed for model networks constructed with the static model and the Erdos-Renyi model while varying the target node fraction.
For the static model, four different power-law exponents are used.
The average degree of each model network is $k_{\text{av}} = 2.5$ and its size is $n=500$.
The input node fraction $n_{\text{d}} = 0.5$, chosen such that the pair $(A,B)$ is controllable.
Further aspects like edge weights and values along the diagonal of the adjacency matrix are discussed in the Methods.
Each set of target nodes is chosen randomly from the nodes in the network.
Each point represents the mean value of the control energy taken over 50 realizations.
The error bars represent one standard deviation.
Note the linear growth of the logarithm of the control energy.
The slopes of these curves are the values of $\eta$ corresponding to each set of parameters.
A linear fit curve is provided in grey.
Also, as $\gamma$ grows, i.e., the scale free models become more homogeneous, the slope approaches that of the Erdos-Renyi model.
(\textbf{b}) The same study as in (\textbf{a}) except that $k_{\text{av}} = 8.0$.
The same behavior is seen but note the difference in scales of the vertical axis.
Each point is the mean over 50 realizations, and error bars represent one standard deviation.
(\textbf{c}) The study in (\textbf{a}) and (\textbf{b}) is performed for more values of $k_{\text{av}}$ and the value of $\eta$ is computed for each curve.}
\end{figure}
%
%% Figure 3
\begin{figure}
\centering
\includegraphics[width=0.5\textwidth]{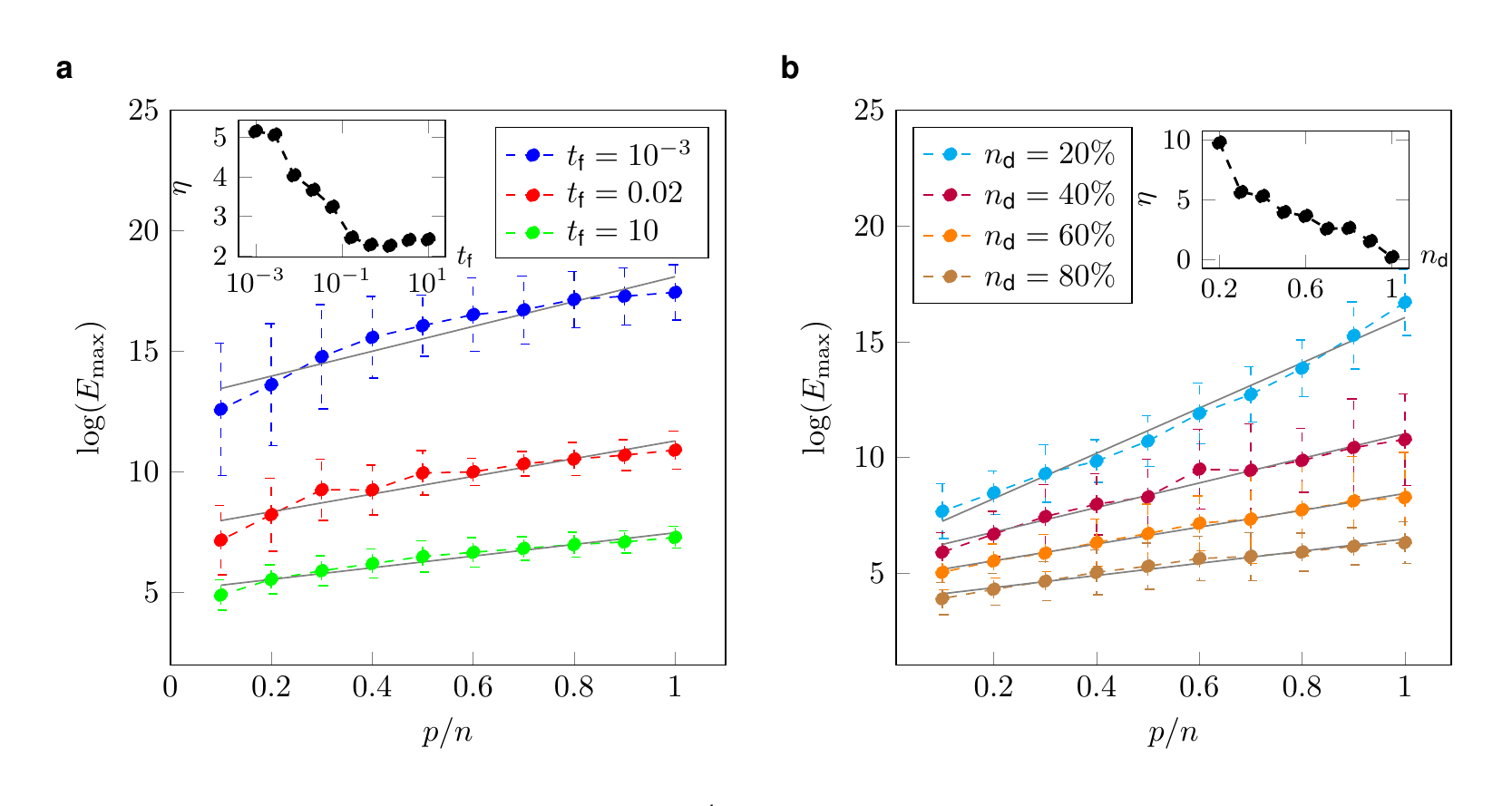}
\caption{
\textbf{Energy scaling as time horizon and input node fraction are varied.}
Besides the average degree and power-law exponent which describe the underlying graph of the network (Fig. 2), there are other parameters that can affect the control energy such as the time horizon and the number of designated input nodes.
(\textbf{a}) The time horizon, defined as $t_{\text{f}} - t_{\text{0}}$, is varied for networks constructed using the static model with the following properties: $n = 500$, $\gamma_{\text{in}} = \gamma_{\text{out}} = 3.0$, $k_{\text{av}} = 5.0$, and $n_{\text{d}} = 0.5$.
As we choose $t_{\text{0}} = 0$, the time horizon is equivalent to just $t_{\text{f}}$.
The main plot shows how the log of the maximum control energy changes with target node fraction, $p/n$.
Each point represents the mean over 50 realizations, and error bars represent one standard deviation.
The inset shows how $\eta$ changes with the time horizon.
We see a sharp increase as the time horizon decreases.
(\textbf{b}) We also investigate how $\eta$ varies with the number of input nodes.
The same class of network is examined as in (\textbf{a}): $n = 500$, $\gamma_{\text{in}} = \gamma_{\text{out}} = 3.0$ and $k_{\text{av}} = 5.0$.
For both simulations, nodes are randomly and independently chosen to be in each target set.
We see that $\eta$ grows as the number of input nodes decreases as shown in the inset.}
\end{figure}
%
%% Figure 4
\begin{figure}
\centering
\includegraphics[width=0.5\textwidth]{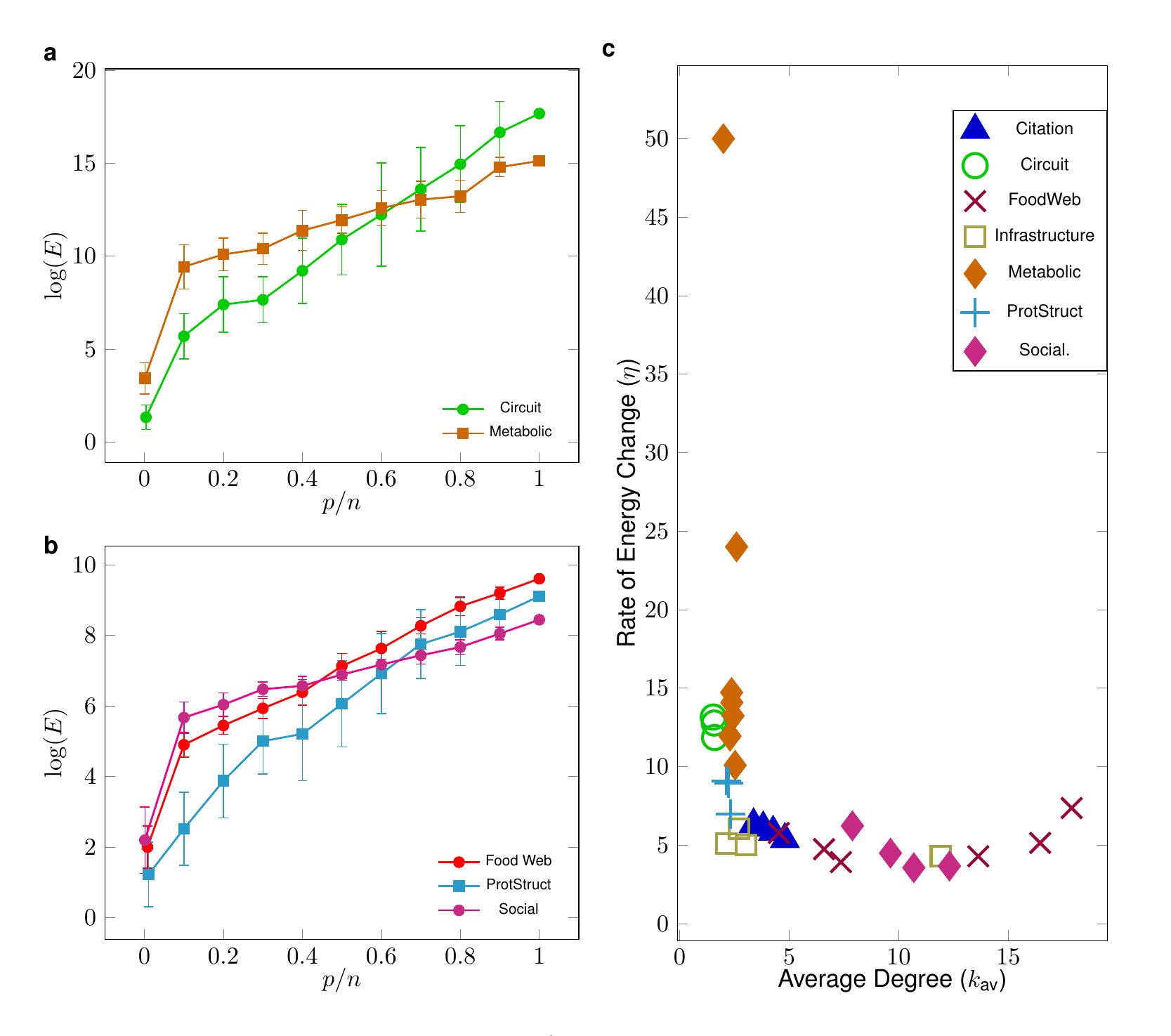}
\caption{
\textbf{Values of $\eta$ for real datasets.}
(\textbf{a}) We compute the maximum control energy required for the s420st circuit network and the TM metabolic network for increasing target node fraction, $p/n$.
Each point represents the mean of fifty realizations where each realization is a specific choice of the nodes in the target node set.
Error bars represent one standard deviation.
(\text{b}) The same analysis performed for the Carpinteria food web, the protein structure 1 network, and a Facebook forum network.
Each points represents the mean of fifty realizations where each realization is a specific choice of the nodes in the target node set.
Error bars represent one standard deviation.
For both (\textbf{a}) and (\textbf{b}), the linear behavior exists only when the target fraction increases greater than $p/n = 0.1$.
(\textbf{c}) We numerically compute values of $\eta$ for real datasets (compiled in Supplementary Table 1) for comparison when $n_\textbf{d} = 0.45$ or larger.
The values of $\eta$ are plotted against each network's average degree as the degree distribution that best describes the degree sequence may or may not be scale-free.
Nonetheless, we see a similar trend, that low average degree networks have a larger value of $\eta$, as demonstrated in Fig. 2(\textbf{c}).
Also worth noting is that networks from the same class (as defined in the legend) tend to have similar values of $\eta$.}
\end{figure}
%
%% Figure 5
\begin{figure}
\centering
\includegraphics[width=0.5\textwidth]{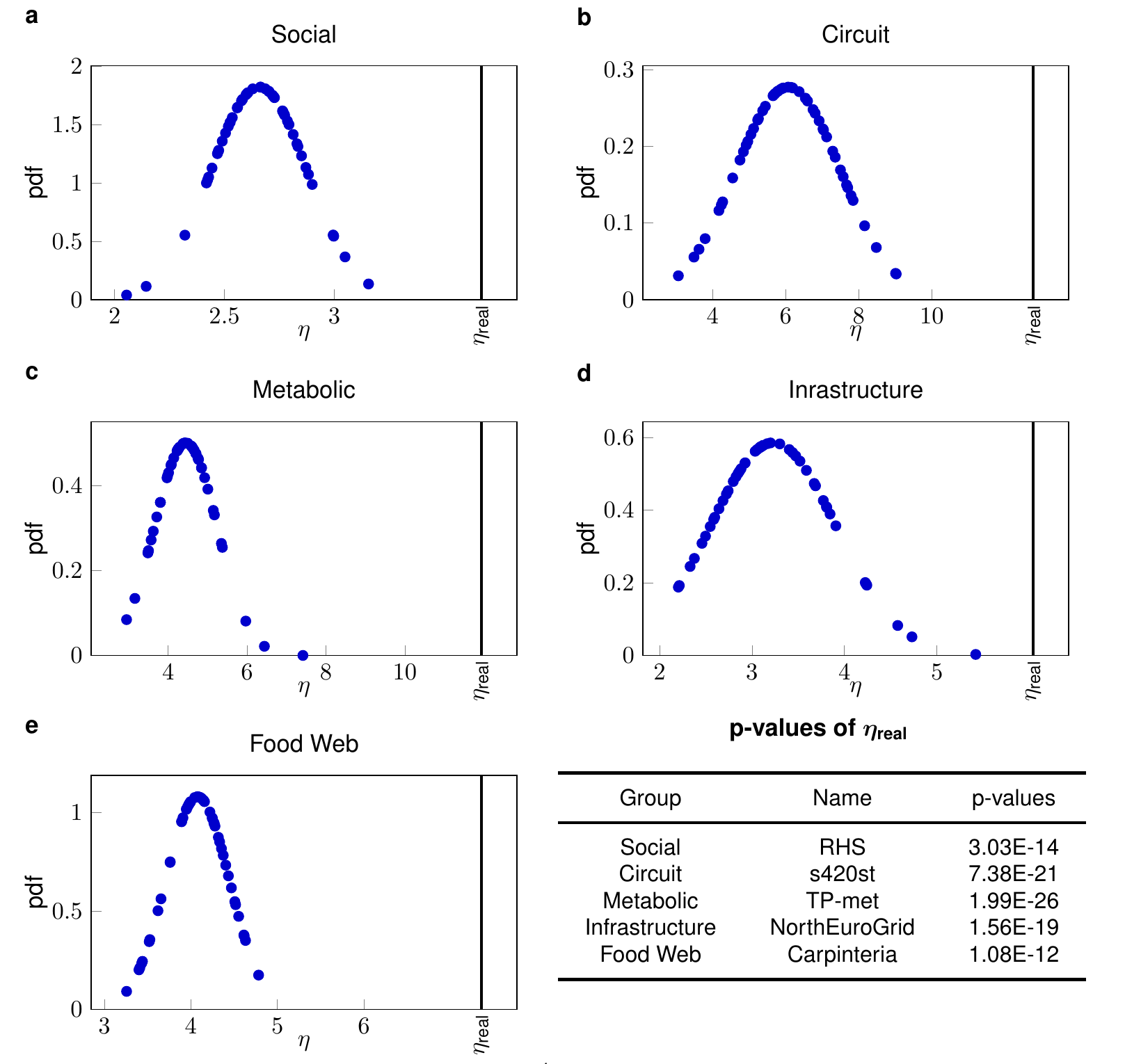}
\caption{
\textbf{Degree-Preserving Randomization of Real Networks.}
Probability density functions (PDF) of the distribution of $\eta$ for a selection of real networks that have undergone degree preserving randomization (DPR) and $n_{\text{d}} = 0.45$.
(\textbf{a}) RHS \cite{freeman1998exploring} from Social.
(\textbf{b}) s420st \cite{milo2004superfamilies} from Circuit.
(\textbf{c}) TP-met \cite{jeong2000large} from Metabolic.
(\textbf{d}) North Euro Grid \cite{menck2014dead} from Infrastructure.
(\textbf{e}) Capinteria \cite{lafferty2006food} from Food Web.
(\textbf{f}) Each of the corresponding p-values are listed in the table.
The vertical lines mark the value of $\eta_{\text{real}}$ which corresponds to the original network.}
\end{figure}
%
%% Figure 6
\begin{figure}
\centering
\includegraphics[width=0.5\textwidth]{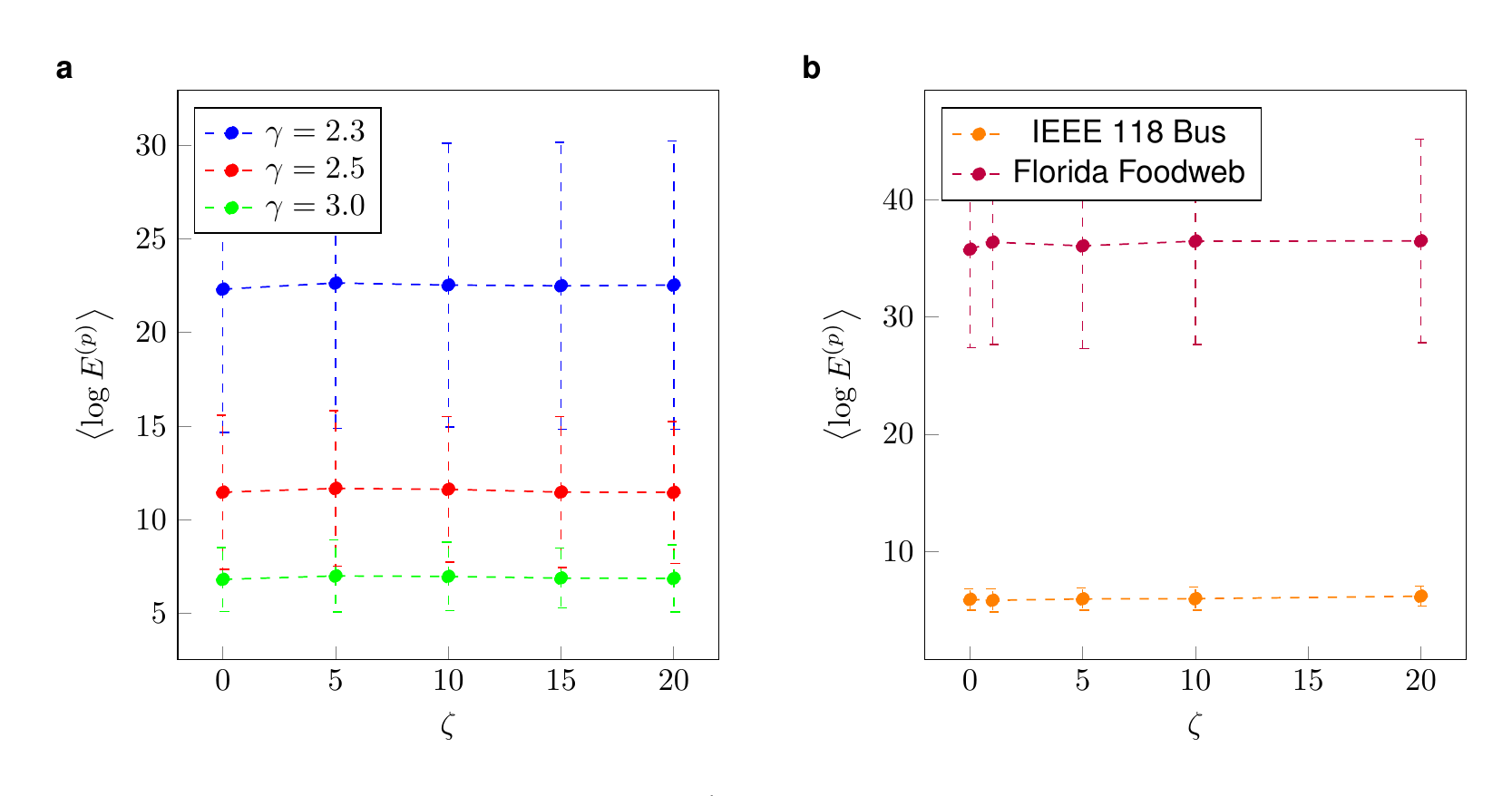}
\caption{
\textbf{Average energy for increasing state weight matrices.}
We demonstrate that for both model networks and real datasets, increasing $\zeta$ (where the state weight matrix, $Q = \zeta I_n$, does not significantly increase the average energy.
(\textbf{a}) The static model is used to generate model networks with parameters $n=300$ and $k_{\text{av}} = 5.0$ where $n_{\text{d}}=0.5$.
Note that the order of magnitude, here represented as a linear scale with respect to the logarithm of the energy, is approximately constant.
Each point is averaged over 50 iterations of model networks and final desired states, which have Euclidean norm equal to one.
(\textbf{b}) Two real networks are also examined and the average energy is computed.
Each point is the mean over 50 realizations where each realization represents a choice of final condition such that the final condition has Euclidean norm equal to one.
For both studies, error bars represent one standard deviation.}
\end{figure}
%
%% Figure 7
\begin{figure}
\centering
\includegraphics[width=0.5\textwidth]{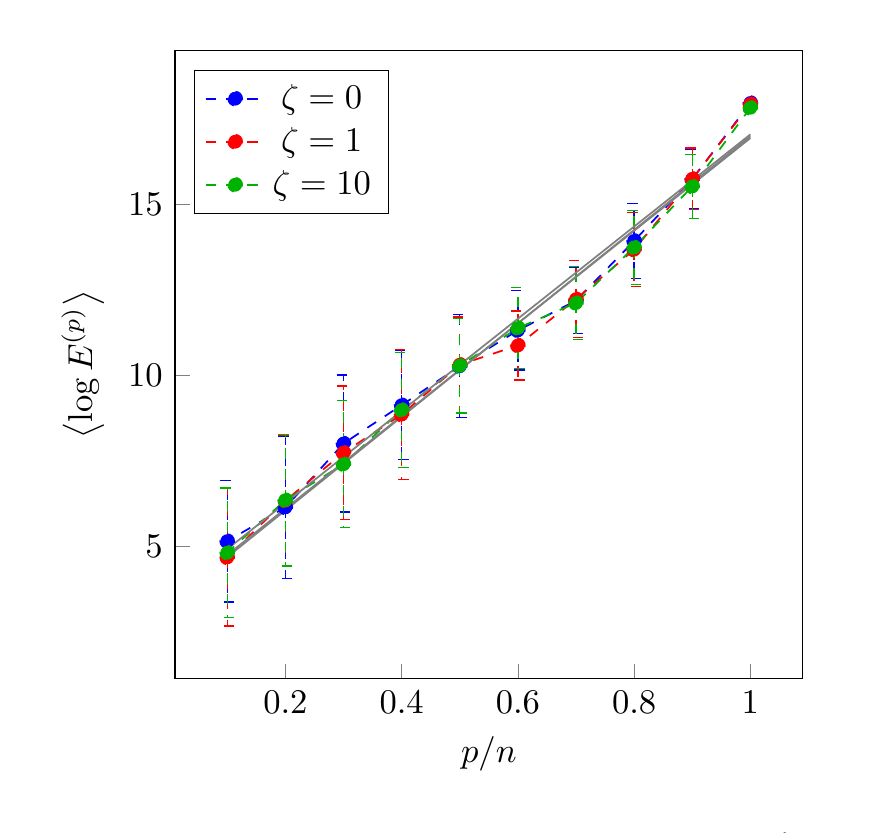}
\caption{
\textbf{Energy reduction for increasing state weight matrices.}
We construct a single model network using the static model \cite{goh2001universal} with the parameters $n = 300$, $n_{\text{d}} = n/4$, $\gamma = 2.7$, and $k_{\text{av}} = 5$.
The energy scaling is examined for the general quadratic cost function.
We compute $\eta$ for different values of $\eta$ such that the state weight matrix $Q = \zeta I_n$.
The values of $\eta$ for $\zeta = 0,1,10$ are $\eta = 13.46,13.66,13.53$, respectively.
Each point is averaged over 50 iterations of target node sets.
The simulations are performed with initial condition set to the origin and the final condition chosen randomly such that $||\textbf{y}_{\text{f}}|| = 1$.
Error bars represent one standard deviation.}
\end{figure}
%
%% End
\end{document}